\documentclass[3p,times,twocolumn]{elsarticle}

\usepackage{ecrc}

\volume{00}

\firstpage{1}

\journalname{Nuclear Physics B Proceedings Supplement}

\runauth{D. Fargion, P. Oliva}

\jid{nuphbp}

\jnltitlelogo{Nuclear Physics B Proceedings Supplement}

\usepackage{amssymb}

\usepackage[figuresright]{rotating}

\begin{document}

\begin{frontmatter}



\dochead{}

\title{Tau Now}

\author[1]{D. Fargion}
\ead{daniele.fargion@roma1.infn.it}
\author[2]{P. Oliva}
\ead{pietro.oliva@unicusano.it}

\address[1]{Physics Department, Rome University 1 and INFN rome1, Ple. A. Moro 2, 00185, Rome, Italy}
\address[2]{Department of Electrical Engineering, Niccol\`o Cusano University, Via Don Carlo Gnocchi 3, 00166 Rome, Italy}

\begin{abstract}
Ultra High Energy Cosmic Rays and UHE neutrinos may lead to a new deep astronomy. However the most recent
results on their correlations and clustering seem to most authors inconclusive. We briefly remind some UHECR models and past and recent results. Our reading and overlapping of IR-gamma-UHECR maps and their correlations seem to answer to several
 key puzzles, offering a first hope of the UHECR astronomy, mostly ruled by lightest nuclei from nearby Universe.
 Regarding the UHE neutrino we recently noted that the flavor ratio and the absence of double bang in IceCube  within highest energetic ten events may suggest still a dominant noisy prompt component. However a first  correlated UHE crossing  muon with expected location (through going upward muon neutrino or horizontally) in IceCube is in our view a milestone in neutrino astronomy road map, possibly partially related, to galactic UHECR narrow clustering. The disturbing and persistent atmospheric neutrino noises,
 both conventional and prompt, call for a better filtered neutrino astronomy: the tau  neutrino ones.
 There are no yet (at present, detectable) TeV-PeVs or more energetic tau neutrino of atmospheric , conventional or prompt nature; only astrophysical ones might soon shine.
 Double bangs in IceCube and in particular the tau air-showers in large array are the unique
 definitive expected signatures of astrophysical signals. In particular tau air-shower amplify in a huge way the otherwise single lepton track, once in decay in flight, into
 a richest three of secondaries (up to a million of billion Cherenkov photons for PeV tau energy) whose wide areas may extend up to nearly kilometer size. Such airshowers are very directional. PeVs energetic tau lepton penetrate hundreds meters inside  the rock before its decay. Therefore horizontal tau air-shower in front of deep, wide valleys or mountain cliff \cite{0e2}, as well as up-going tau air showers escaping our Earth, observable in air by their fluorescence lights as  in AUGER and  TA,  might be a signal at EeVs energies. At lower energies  blazing few  PeVs tau airshower flashes are better observable from the top of the mountains, by an array located in a crown edge (as water Cherenkov, telescope Cherenkov or radio array), as it has been done in ASHRA or it might be done  on ideal modified GRAND experiments constructed within aeolian towers by radio array possibly in mountains, facing the wide peculiar $\tau$ neutrino sky: our own Earth.

\end{abstract}


\end{frontmatter}

\section{Introduction: Nine astronomy at birth}
\label{sec:intro}
Several new astronomy are beyond the corner: at least nine way to see the sky.
Astronomy, as it sound the name, it is somehow the way to give a name to an astro; indeed in the past we begun from constellations maps to recognize the sources and to describe it in the correlated sky map.
The most popular and old astronomy is the optical one. Since  Galileo we did explore the sky better and better
with larger and larger telescopes in visible lights since four centuries.
Last century, mostly thanks to Maxwell, we explored the sky in much different and surprising colors:
 from radio wave-lengths to infrared, from ultraviolet to X and hard gamma sky. At each energy windows we find novel
  often unexpected  actions and actors. The discover of  Cosmic Rays (CR), since last century, being made of  charged particles, has  not led to any astronomy because charged CR are bent by galactic and cosmic magnetic fields whose bending makes CR directionality  on Earth lost and a smeared cloudy noise almost at all energies. Large scale magnetic (and not electric) fields exist because no magnetic monopoles charge have been left by Nature in our universe. However the highest CR, above tens EeV, they might be able to survive the Lorentz bending offering a directional astronomy.
 Moreover since half a century, as soon as Penzias and Wilson discovered the cosmic big bang noise, it has been noted that these soft infrared relic thermal photon
  might appear to UHECR in flight, in their boosted reference system, as hard gamma as, via resonant $\Delta^{+}$ photopion production, to slow and stop nucleons
  making UHECR  flight  bounded in a GZK narrow Universe \cite{0a, 0b}; because of such a GZK cut off the UHECR Universe is as small as $\simeq1\%$ the cosmic size.
  Therefore very easily to be correlated with a million part cosmic source volumes.
  \subsection{UHE EeV galactic neutrons}
  At EeV energies peculiar UHECR, made by neutrons may survive within our galaxy scales offering a well directional astronomy.
  Unfortunately not yet observed. A much higher statistical signal are needed. Indeed even a more abundant PeV CR flux is not clustering
  and it is not shining yet PeV gamma astronomy
  that it is  bounded, by gamma-cosmic background radiation opacity, in similar galactic  volumes.
  No point source, but just large scale anisotropy has been revealed by PeV IceCube downward muons.
 \subsection{Six astronomy at horizons}
  Out of UHECR and EeVs neutrons astronomy  six more astronomy are coming to birth: the three flavor neutrino and anti-neutrino ones.
  The well known electron neutrino shined from the Sun as well as from the SN 1987A rare one. Two historical well tested astronomy.
   Additional, but random, relic supernova signals might be soon revealed by SK revived with gadolinium.
   However at higher GeVs TeVs  energies,  abundant atmospheric neutrinos as being the secondary parasite of bent cosmic rays, are hiding the (otherwise expected) astrophysical signals. Only at ultra high energy  the (expected)  harder astrophysical signals might overcome the softer atmospheric ones: indeed since four years  at $60$ TeVs $-$ PeVs energy  the sudden neutrino flavor change in IceCube had hint for a new windows to astrophysical  windows whose  directionality might be better led by $\nu_{\mu}$, $\overline{\nu}_{\mu}$  tracks in underground km$^{3}$ IceCube detector. The new dominant showering events are made by CC $\nu_{e}$, $\overline{\nu}_{e}$,  by  CC $\nu_{\tau}$, $\overline{\nu}_{\tau}$  as well as by all NC $\nu$, $\overline{\nu}$; they are in IceCube ice, smeared in wide $\simeq 10^{\circ}$ spot making a poor
  smeared astronomy at the moment.
  Not yet sharp correlation with these UHE neutrino and known gamma sources is (apparently) available.
  A peculiar $\overline{\nu}_{e}$  Glashow resonant signature at $E_{\bar{\nu_{e}}} \simeq 6.3$ PeV is still absent. Future  $\nu_{\tau}$ double bang, due to its birth and decay, in IceCube or the competitive tau airshower born on Earth ground and shining in the sky by fluorescence, Cherenkov, X , gamma and radio bangs \cite{0e1, 0e3, 20}.
  The very recent (and promising) correlation of a rare AGN flaring activity (a major outburst during 2012 several months of the blazar PKS B1424-418 \cite{001}) within the largest PeV neutrino shower event (n. $35$) in ICECUBE is nevertheless exciting but still questionable because of the  wide arrival solid angle of any neutrino shower and because of the long time span windows considered.

\section{A ninth gravitational Waves astronomy?}

  The ninth and most exciting (in particular by rumors just on the web these days)  astronomy, to be published this week, the Gravitational Wave one, GW, seem in our view too much difficult to be correlated (by single timing array triangulation) to any far source. The Earth size is so small and the millisecond scale time is so wide that only too wide solid angle might be pointed out. In analogy let's remind the inability of widest satellite distances (as Compton, HETE-2, SWIFT, INTEGRAL, RXTE, ULYSSES network) to identify by triangulation themselves the GRBs location in last half a century: only Beppo-Sax by its X-ray tracking was able to point out the galaxy leading to the optical afterglow discover and to the cosmic GRB identification.
  Moreover a collapse of a binary Black Hole without any accretion disks around, even of several tens of solar mass  will make (as long as we know) no associated lightening in photon (or neutrino) in the sky. Therefore we are afraid that any millisecond  timing of a probable discovered black holes binary collapse (offered by LIGO, north of Richland and  LIGO in Louisiana) will mark the event  only  within a very  disk area in the sky, with no correlated source, just for a statistical cosmic counting and estimate.  On the contrary as we will argue, the horizontal upward UHE crossing muons \cite{0c} in ICECUBE, or any double bang and in particular the tau air-showers \cite{0e2, 0e1, 0e3} traces might lead soon to PeVs up to  EeVs signals by their sharper and better directionality.
 \section{More astronomy at the horizons?}
  The nine GW astronomy list might be extended to a wider (and more hidden) one,  containing possible (but yet unobserved) Susy UHE signal. For instance the most on fashion
  tenth astronomy  might be indebt to UHE PeVs neutralino $\chi$ whose eventual production would be hidden in deep energetic AGN or in micro-quasars jets and  whose  signature would be complementary to the Glashow $\bar{\nu_{e}} + e\mapsto W^{-}$ resonant shower: the $\chi$ electron production of selectron  $\chi + e \mapsto \tilde{e}$  resonance would  showers in ICECUBE \cite{002} in a very similar ways as the expected Glashow one.

\section{UHECR maps and bending}

The high-energy cosmic radiation known as Cosmic Rays (CR) is fundamentally made up by electric charged nuclei and because of this the traveling particles that constitute CR are deflected by galactic and extragalactic magnetic fields, resulting in a smeared map of their arrival direction at Earth. Since the discovery of CR, more than a century ago,  due to the magnetic fields deflections no certain sources nor clear correlation with other astrophysical maps are discovered yet. As of today it is not even crystal-clear if CR origin is to be ascribed to a galactic, extragalactic or even to a cosmic relic nature, if not to a mix of them.  Moreover the same CR source might be able to produce Pions or Kaons or charmed mesons whose final signals will shine at highest energies, overcoming atmospheric noises, as  astrophysical ones.
\begin{figure}[t]
    \includegraphics[width=0.53\textwidth]{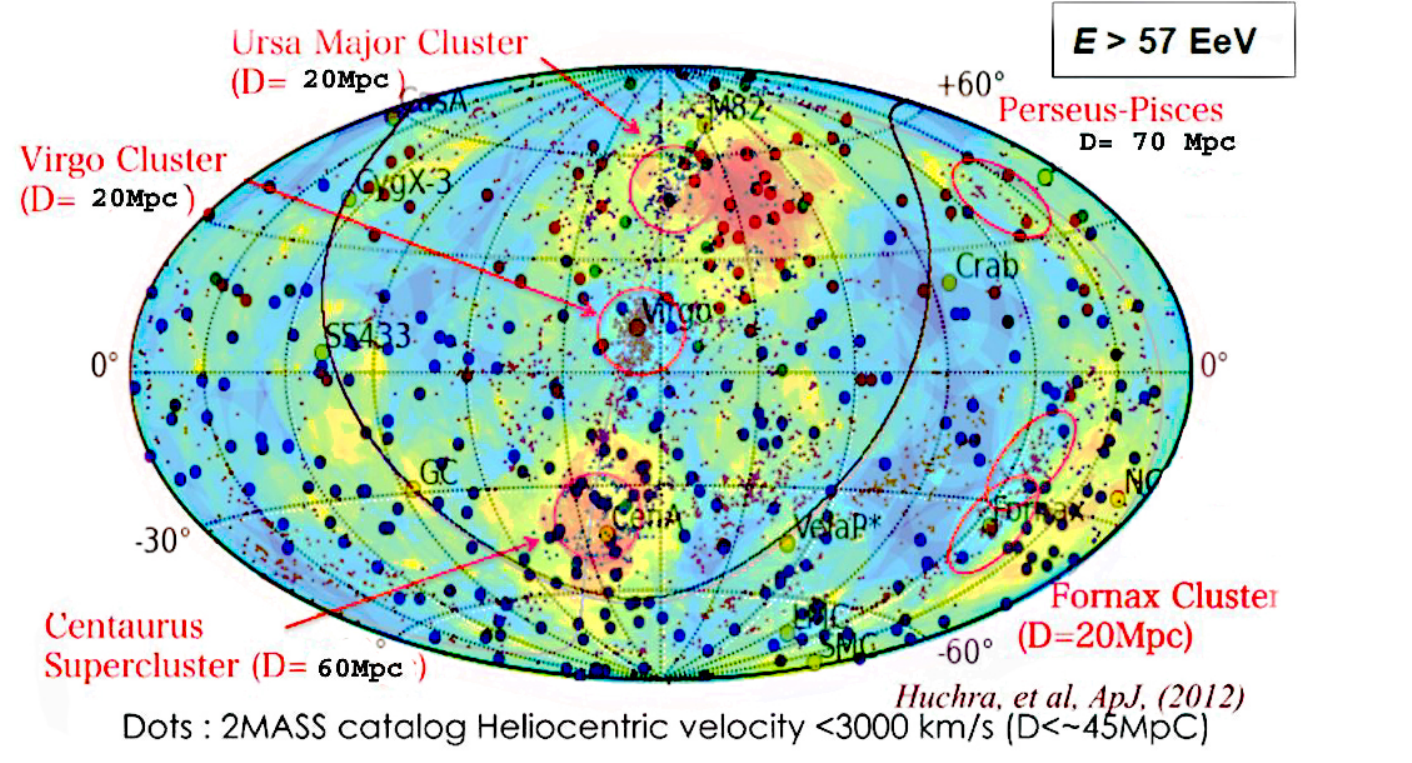}
    \caption{UHECR event map (TA as red dots, AUGER as blue ones) over the nearest Galaxy mass density within GZK.
    The absence in the main map center of nearby Virgo cluster where most
    sources of UHECR might be expected, it is very surprising and remarkable. The presence of two main
    colored, red, anisotropic clustering, the North and South Hot Spot, are the main shining large $10^{\circ}-20^{\circ}$ anisotropy; narrow clustering are also shown \cite{6b, 8}.}\label{fig1}
\end{figure}

\begin{figure}[t]
    \includegraphics[width=0.5\textwidth]{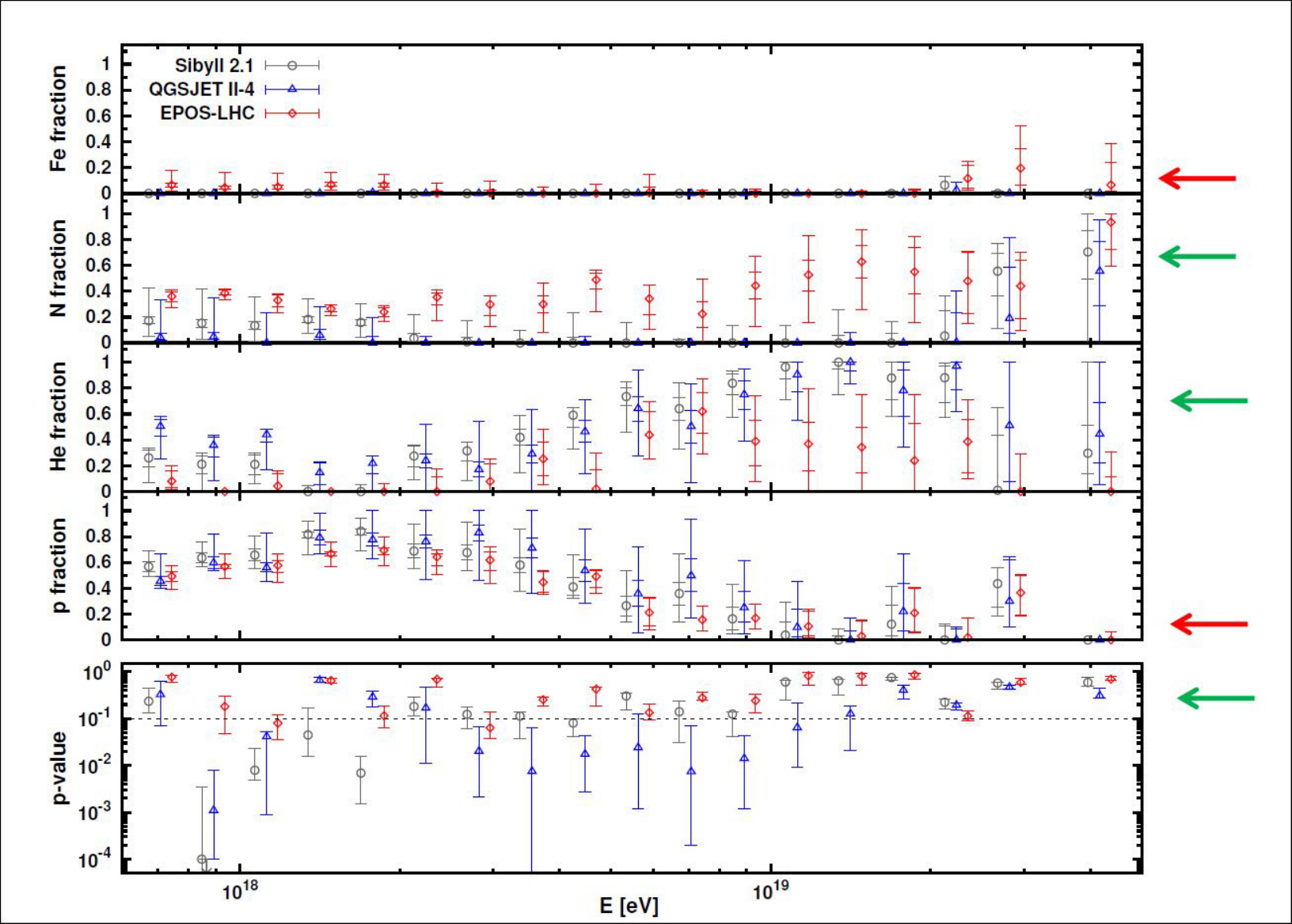}
    \caption{Recent composition model able to fit the UHECR airshower profile. The light and lightest nuclei
    are the only ones able to fit at best the data \cite{9}; the same composition has been proposed  long before
    by much earlier articles \cite{6, 6b, 8}.}\label{fig02}
\end{figure}

Apparently no guaranteed connection up to now \cite{1} has been found even between High-Energy $\gamma$ rays sources and Ultra-High-Energy CR (UHECR) whose magnetic rigidity allows them in principle to reach us un-deflected, pointing then back to their sources. Their deflection maybe coherent or random \cite{6d}; for the latter random one as large as
$$
\delta\theta\simeq8^{\circ}\cdot\; Z \; \frac{6 \cdot 10^{19}\,\mathrm{eV}}{E_{\mathrm{UHECR}}}\frac{B_{\mathrm{Gal}}}{3\,\mu\mathrm{G}}\cdot\sqrt{\frac{L}{20\,\mathrm{kpc}}}\cdot\sqrt{\frac{l_{\mathrm{Coherence}}}{2\,\mathrm{kpc}}}
$$
For $Z= 2$ or $Z= 4$  as for Helium or Beryllium, this deflection angles \cite{6d} are comparable with observed smeared UHECR clustering observed  see Fig.\ref{fig1} in last years (the Hot Spot), a composition to be discussed below; see Fig.\ref{fig02}.

 \section{The scandal of the Virgo absence in UHECR sky}

 Moreover, UHECR being bounded ($D_{\mathrm{GZK}}\sim$ 50~Mpc) by photo-pion opacity (if UHECR are protons) or by much narrow photo nuclear dissociation distances  ($D_{\mathrm{GZK}}\sim$ 3 Mpc), if UHECR are light nuclei, they are theoretically doomed to be confined in a very local  (or nearby) universe and because of this they should be better display a correlation with known galactic, local group or near cluster of galaxy sources. Virgo cluster, as shown in Fig. \ref{fig1} center, is missing.

\subsection{UHECR above GZK and relic neutrinos}

In the last two decades a huge effort of experiments (Fly's Eye, AGASA, HiRes, Auger, TA), aimed to reveal such powerful (but rare) UHECR sources, by collecting hundreds of such signals. On 1995 a first (most powerful up-to-date) event UHECR discovered by Fly's Eye did show a direction uncorrelated to any nearby source (GZK cut off $D_{GZK}\sim 50 Mpc$). Also AGASA (1990-2000) a Japanese experiment, it did claim as well  the absence of a GZK for uncorrelated events;  we then suggested  that a relic neutrino with mass, spread in dark halo could be a calorimeter where UHE$\nu$ at ZeV energies (originated at cosmic edges by AGN or GRBs) might arrive and hit such relic cosmic $\bar{\nu}$ at rest leading to Z boson resonance  $E_{\nu} \simeq{\frac{M_{Z}^{2}}{2 \cdot m_{\nu}}}$; the Z-boson prompt decay in-flight may shine fragments like $p$, $\bar{p}$, $n$, $\bar{n}$, that might be finally detected on Earth as the UHECR signals, reaching us,  by the help of such transparent ZeV $\nu$ couriers, from any corner of the universe (the so called Z-burst or Z-resonant model \cite{2, 3a, 3b, 3c}).

This model since 1997 till 2007 has been  very popular in solving the apparent absence of a GZK cut off for most energetic Fly's Eye (and AGASA) UHECR event. During that two decade the experimental neutrino mass faded away from earliest ($10-5$ eV) values to a more  light allowed value ($1-0.4$ eV) or even just as tiny as the lightest allowed mass comparable to the atmospheric neutrinos mass splitting ($\triangle m_{\mathrm{atm}}\sim 0.05$ eV). The very tuned value ($0.4$ eV) mass was found an ideal one to feed UHECR as large as $\simeq 2 \cdot 10^{20}$ eV \cite{3a, 3c}.

 \subsection{Sterile relic neutrinos now?}

 The peculiar tuned neutrino mass to fit the  $2\cdot10^{20}$ eV suggested \cite{2},\cite{3a}, \cite{3c}, is partially successful in gravitational clustering; a slightly larger mass as  $m_\nu\geq1-4$ eV may do better but it will feed mostly UHECR at few tens EeV. Indeed the ten EeV UHECR energy maps are quite  isotropic and smoothed, possibly needing for homogeneity a cosmological connection as the one at tens EeV UHECR events.
  It is remarkable to mention that many (but often somehow controversial) and very recent results may even favorite this sterile neutrino role exactly in the required mass range: $\delta^{2} m_{\nu} \simeq2$ eV$^{2}$ \cite{3e}.
  However since 2001 the Hires UHECR detector by fluorescence telescope array and later the largest area array as AUGER (and more recently the Telescope Array, TA), apparently, finally found an UHECR spectra cut-off possibly related to the GZK one. Some Hot spot large scale anisotropy are found, uncorrelated with cosmic homogeneity. Therefore there is, once again apparently, no longer need for a tens ZeV cosmic neutrino relic neutrino connection at highest energies; nevertheless  we may keep in mind that the tens EeV UHECR  homogeneous maps may  appeal, if really statistically probed, to a cosmic sterile neutrino Z burst solution.

 \subsection{The UHECR Science 2007 correlation}

Indeed on November 2007 the Auger Collaboration, based on earliest 26 UHECR events claimed that those events anisotropy were correlated with the Super Galactic Plane \cite{4}, in complete agreement with the GZK cut off  expected Universe size. Unfortunately or more puzzling, as it appeared in the same article, there were two evident  (at least to someone of the community \cite{6}) contradictions: the first being that the UHECR composition imprint (as A. Watson defended it in public \cite{5}) at highest energies were pointing toward heavy nuclei-kind one not to a nucleon one. At that time our argumentation was that such a UHECR nuclei, if as heavy as Fe, would have been severely bent ($\sim90^\circ$), making any correlation with the Super Galactic Plane irrelevant and meaningless  \cite{6}. We then noticed that, if UHECR would be on the contrary lightest nuclei, such as He, Be, B, Li, etc. then the bending wouldn't have been so large ($\sim10^\circ$-- $20^\circ$) \cite{6d}; nonetheless such light \cite{6} or lightest \cite{6b} nuclei have a different remarkable behavior.

\section{Lightest Nuclei for Cen A, M82, Virgo}

The possibility that UHECR are mostly light nuclei \cite{6, 6b} was (and somehow remain, silently) so unpopular that it has not be taken seriously by Auger (or TA) collaboration; indeed the common composition models considered by most authors along last years $2007-2014$ were a mix of nucleon $p$ and iron $Fe$.
The most recent (2015) AUGER composition model and articles \cite{9}, see Fig. \ref{fig2}, had anyway to converge to light and lightest UHECR model; nevertheless for our understanding light nuclei as UHECR was and still is the best hypothesis
 a solution to the second main puzzle hidden in \cite{4}, still neglected  and unsolved: the mentioned scandalous Virgo cluster absence \cite{6}.
We wish to call the dearth of Virgo cluster  for a decade of UHECR events ``a scandal''  because it represents a loud silence in the maps still today, just comparable to the same silence that  most author deserved to this problem.
Such a \emph{non-signal} problem has to be taken into account in particular by all the UHECR \emph{proton} proponents, not a few at the moment,  keeping in mind that in the infrared (within the GZK bounded) sky, Virgo shine as a leading hot spot.

\subsection{The UHECR Hot Spot in North and South sky}

The light nuclei composition for UHECR \cite{6, 6b}, would solve at once this tremendous puzzle: UHECR didn't arrive from Virgo $D_{\mathrm{Virgo}} = 20$ Mpc because the fragile lightest UHECR nuclei, He-like, will be soon fragmented after a few Mpc flight.
Moreover the expected bending and the observed characteristic bending angle of UHECR in smeared
``Hot Spot'' around the nearest Cen A  AGN \cite{6, 6b, 6d}, the nearest and  brightest in gamma and radio AGN  candidate source, whose signals may reach us also for lightest nuclei bounded sky. Indeed  Auger in the last years (2007- 2015) found a  Hot spot around Cen A and in last couple of years TA too found it (where the given name was ``North Hot Spot'' to be distinguished from the Auger ``South one'') \cite{4, 7, 8}. None of these two main event clustering are centered, in fact, on Virgo which is, just to remind it, the nearest, IR brightest of the sky, associate to most large mass cluster within 40 Mpc (main GZK cut off distance for protons), see Fig. \ref{fig1}: we remark the Virgo role because any proton-composed UHECR should shine also or mostly from it; indeed  not by chance, Virgo, is the Italian gravitational wave collaboration name, whose very near prospect was and it is to reveal gravitational waves exactly from nearest Virgo cluster, within the detector threshold.
The North Hot spot UHECR cluster might be correlated  assuming a random magnetic field toward Cen A, \cite{6f, 6d} and, assuming an asymmetric coherent UHECR bending \cite{6d, 17} to M82 \cite{6d, 8},  a second  bright AGN within a nearby (few Mpc) distance, as Cen A \cite{6b}. Also dwarf galaxy Fornax, NG 253,  have been considered possible sources of UHECR \cite{17}.

\begin{figure}[t]
    \includegraphics[width=0.5\textwidth]{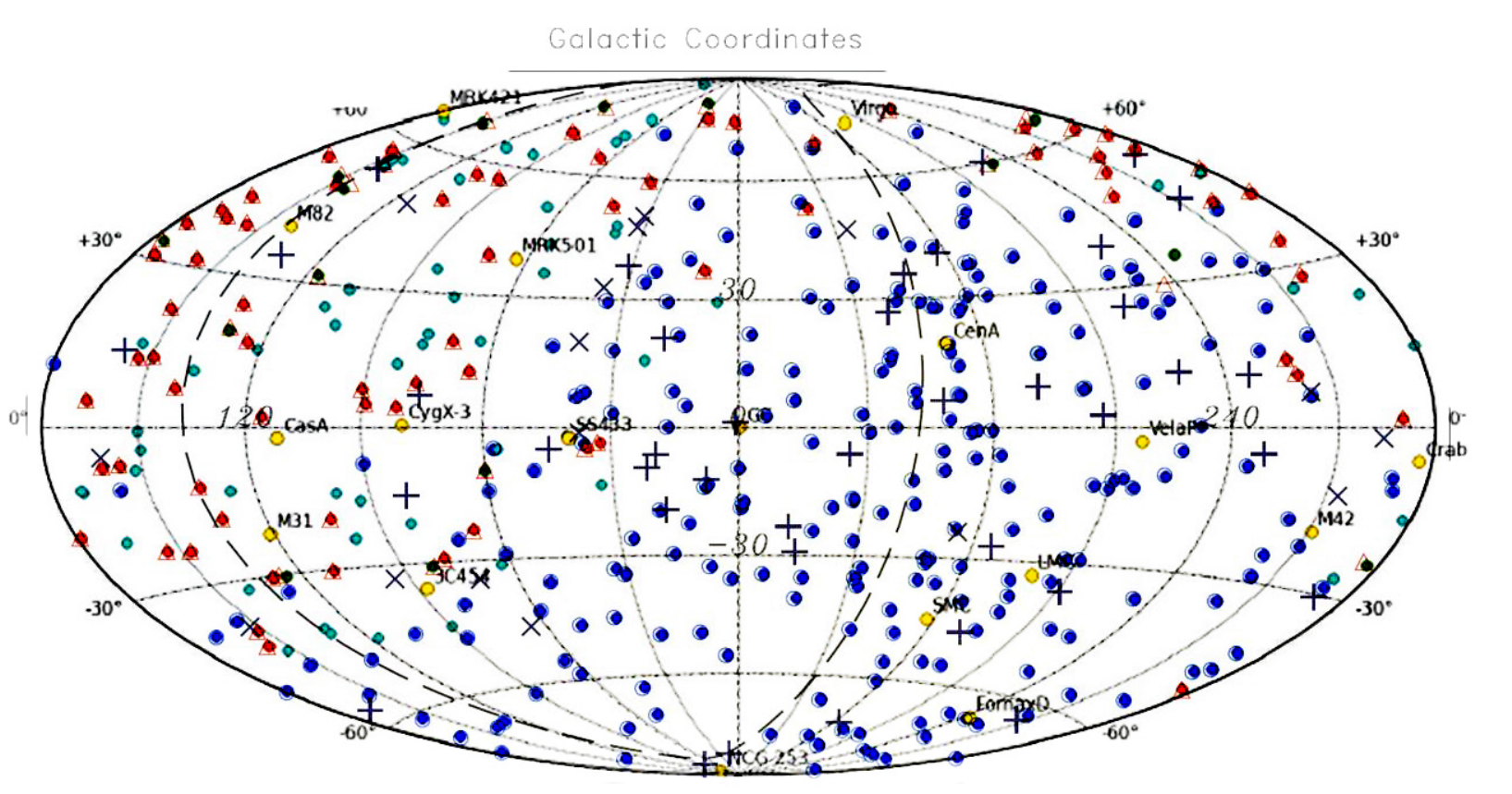}
    \caption{An Earlier UHECR  event map (blue dots, AUGER, while red ones are TA) in galactic coordinate \cite{8} well  overlapped   (as obvious) on  later and more recent   identical one (2016) \cite{1}.  Our additional events (cyan dots) remind AGASA and minor old Fly's Eye event \cite{8}.   The crosses X stand for UHE muon aligned tracks made by hundreds
    TeV  events (7 included and 9 crossing the IceCube detector) in last years; the symbol "+" stand for  neutrino shower at those highest energies whose directionality is much less secure \cite{1}.  A more wide sample of crossing muons is considered in next pictures.}\label{fig00}
\end{figure}

\begin{figure}[t]
    \includegraphics[width=0.5\textwidth]{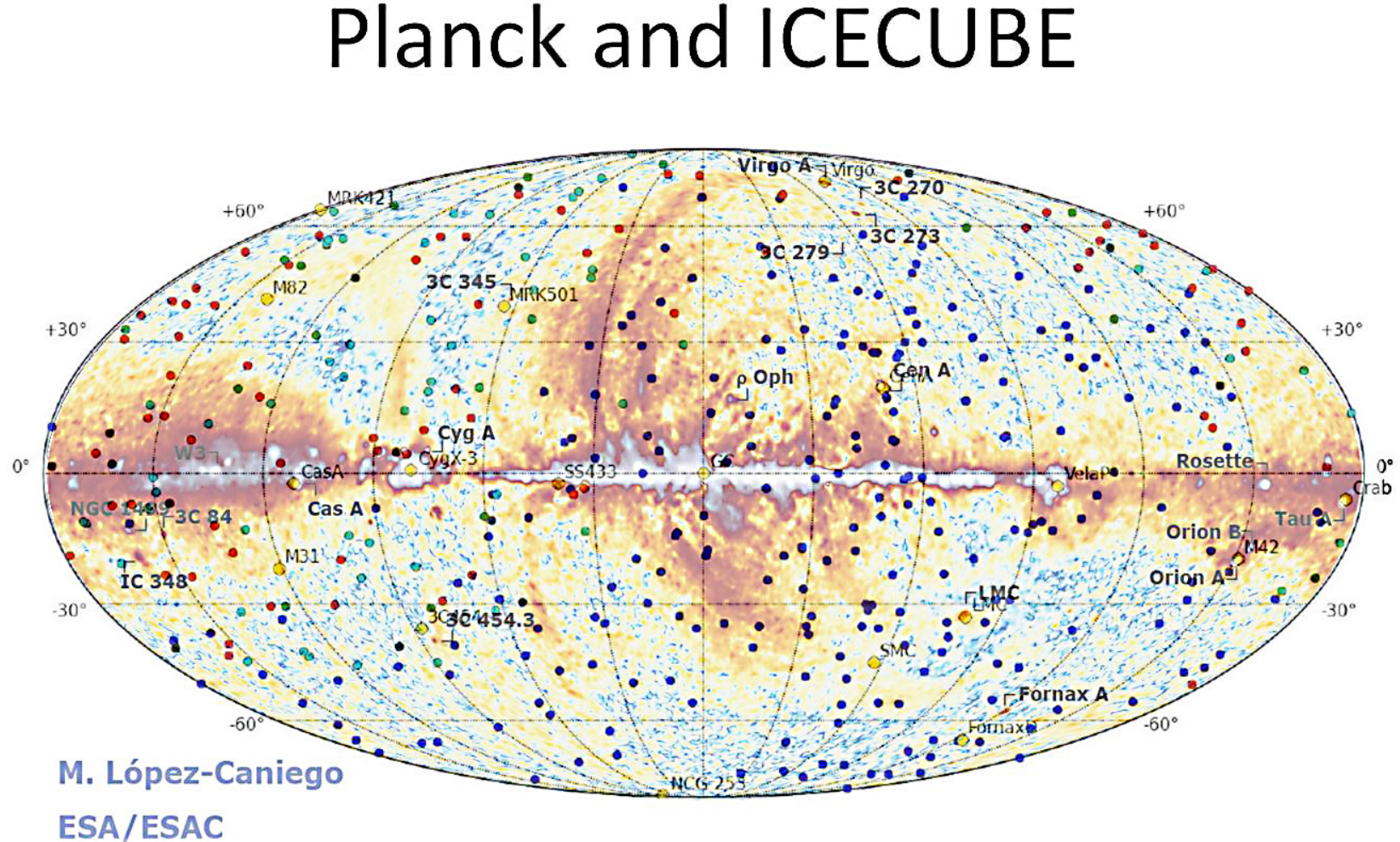}
    \caption{\label{fig2}  UHECR events over last Planck IR sources. Note the apparent clustering of the UHECR along the boundary of the  IR Planck galactic map. Note also the clustering around the Vela source, SS433, LMC, Cyg X3 \cite{8}. }\label{fig04}
\end{figure}

\begin{figure}[t]
    \includegraphics[width=0.48\textwidth]{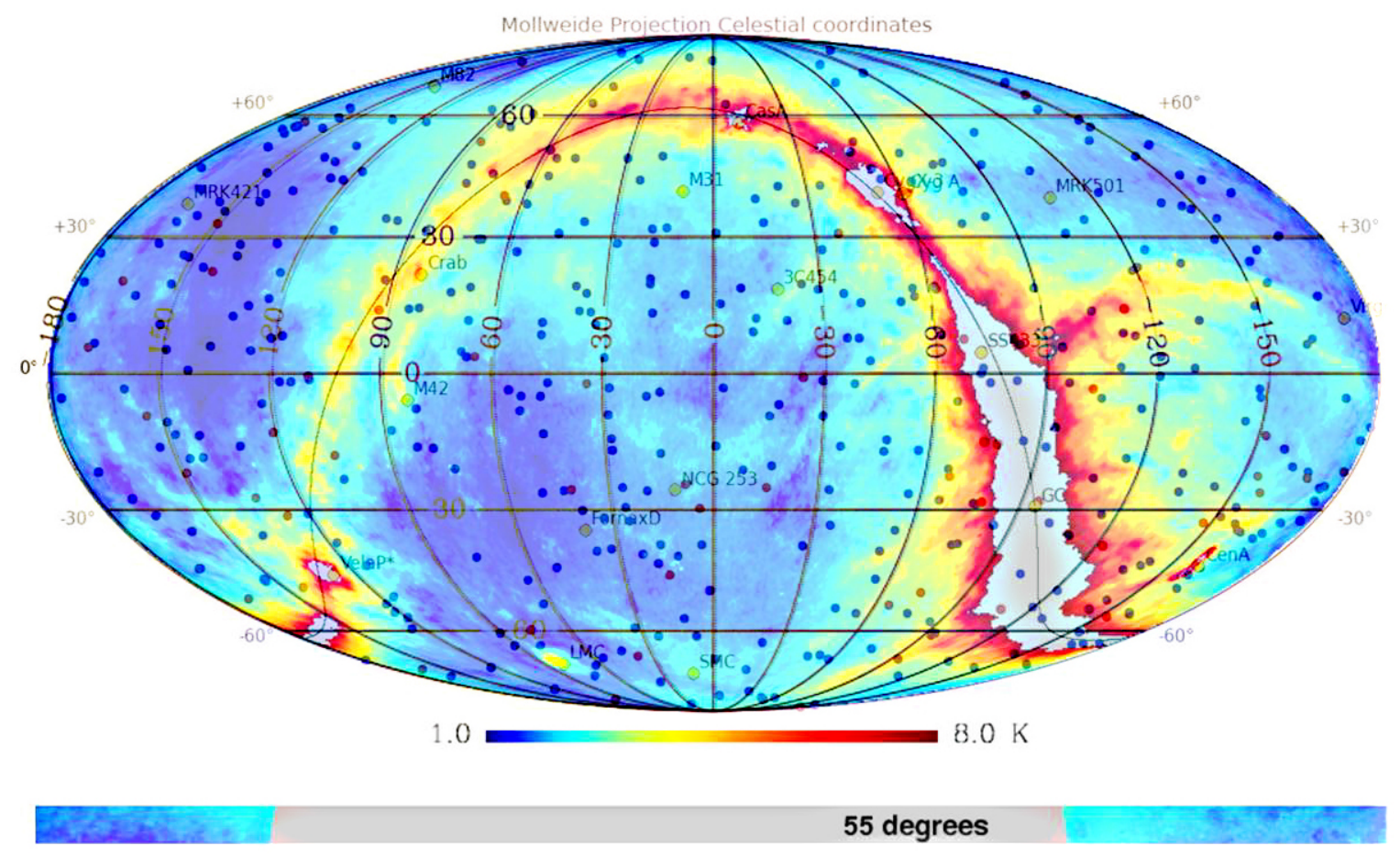}
    \caption{UHECR in anti-centrum celestial coordinate and IR sky. Note the paucity of the events in the same galactic IR plane versus the large abundance of the events around the same IR galactic plane boundary.
    The higher galactic magnetic field density along the galactic spiral center, the higher bending and screening ability of these fields is the possible cause of it.}\label{fig05}
\end{figure}

\subsection{UHECR on the galactic ridge ? }
The rapid inspection of the above UHECR in galactic or celestial coordinate UHECR distribution Fig. \ref{fig04},\ref{fig05}  do show a characteristic signature:
the paucity of UHECR events at the Galactic center, at Vela center, at LMC area and in general along the same galactic plane.
There is also a denser UHECR presence along the galactic ridge. Therefore this signature, if not just a miss-leading statistical fluctuation,
it may imply a shielding and a screening  of the galactic arm magnetic fields (as well as of the local Vela, LMC fields). It also
favor a possible location of UHECR sources ( possibly binary system) escaping from galactic plane in those area. The future maps might confirm or
refute this preliminary suggestion.

\section{The Composition Solving the Virgo scandal}

The lighter nuclei as $^4$He are too fragile to fly inside the BBR more than a few Mpc  \cite{6, 6b} resulting in no shining and screening  Virgo cluster,  solving then its embarrassing absence. This simple screening of UHECR from Virgo (by lightest nuclei composition)  did inspired us and made us accept the UHECR Auger slant depth message:
UHECR composition had to change from a proton-like nucleon at EeVs up to ten EeV energies toward heavier nuclei such as He, Be, B and so on. While for most authors during  last years a proton-iron mixture has been the ideal model, only recent Auger studies  of air-shower profiles \cite{9, 10} of UHECR at few tens EeV confirmed that not  proton nor iron is  fit to observations: only light (N nitrogen) up to lightest (He) nuclei are the best candidates for UHECR. \cite{11, 12}. Our simplest assumption of light nuclei explained the UHECR cluster of events around Cen A. In this view we  argued and foresaw \cite{13, 14, 15} that fragments of this light nuclei from Cen A must be present at lower energies (D, $^3$He, p) tracing, like an elonged tail, the same UHECR bundle of events along with Cen A. These predictions were published a couple of years before the Auger observations of 2011. In the latter years the possible clustering along Vela, Cyg-X3 and maybe Crab and Large Magellanic Clouds have been enriched with a very rare highest UHECR overlap between Auger and TA events, as reported in \cite{12}; this location is along the galactic plane pointing toward SS433, a well known galactic binary precessing micro-jet. The source SS433 location had additional correlated UHECR  events by Auger so that the multiplet is now a very promising UHECR source candidate. Indeed as shown in Fig. \ref{fig03}-\ref{fig04}, one of the last (21) UHE through going muon by UHE neutrino in IceCube is coincident with the UHECR narrow events cluster.

In the period 2013 -- 2014 we noted that a similar triplet, near the galactic plane,
is correlated with an old HiRes event and Auger event number 5, making this particular location a serious source candidate \cite{16, 17}.

\section{ The discover of first astrophysical $\nu_{\mu}$ sources?}

On 2013 we pointed out that UHECR might be correlated with UHE$\nu$ by IceCube. However, while UHECR shower in IceCube arrive at angles of $\sim\pm15^\circ$, UHECR muon track offers a much narrow direction $\sim\pm0.5^\circ$, therefore we claimed and foreseen that muon tracks (both born inside or outside IceCube) that are able to cross the detector (also called trough-going) have to be the cornerstone of UHE$\nu$ astronomy \cite{18}. In this period of random rush for the source
it must be kept in mind the message: track are the only ones, today to offer a sharp test of muon neutrino astronomy.
We estimated that few dozens of future events could open such a new astronomy; our predictions were followed by later article considering UHECR candidate clusters, four labeled galactic sources (Cyg-X3, SS433, Vela and UHECR event n.5 and Cen A, our unique nearby AGN and the brightest in X-ray sky) \cite{8a}; see also \cite{8}. We wrote that paper \cite{8a} two years before IceCube disclosed just few months ago its largest sample of 21 crossing muon, discussed and overlap to our earlier sources in figure below.

\subsection{The Crossing muons richest maps}

\begin{figure}[t]
    \includegraphics[width=0.5\textwidth]{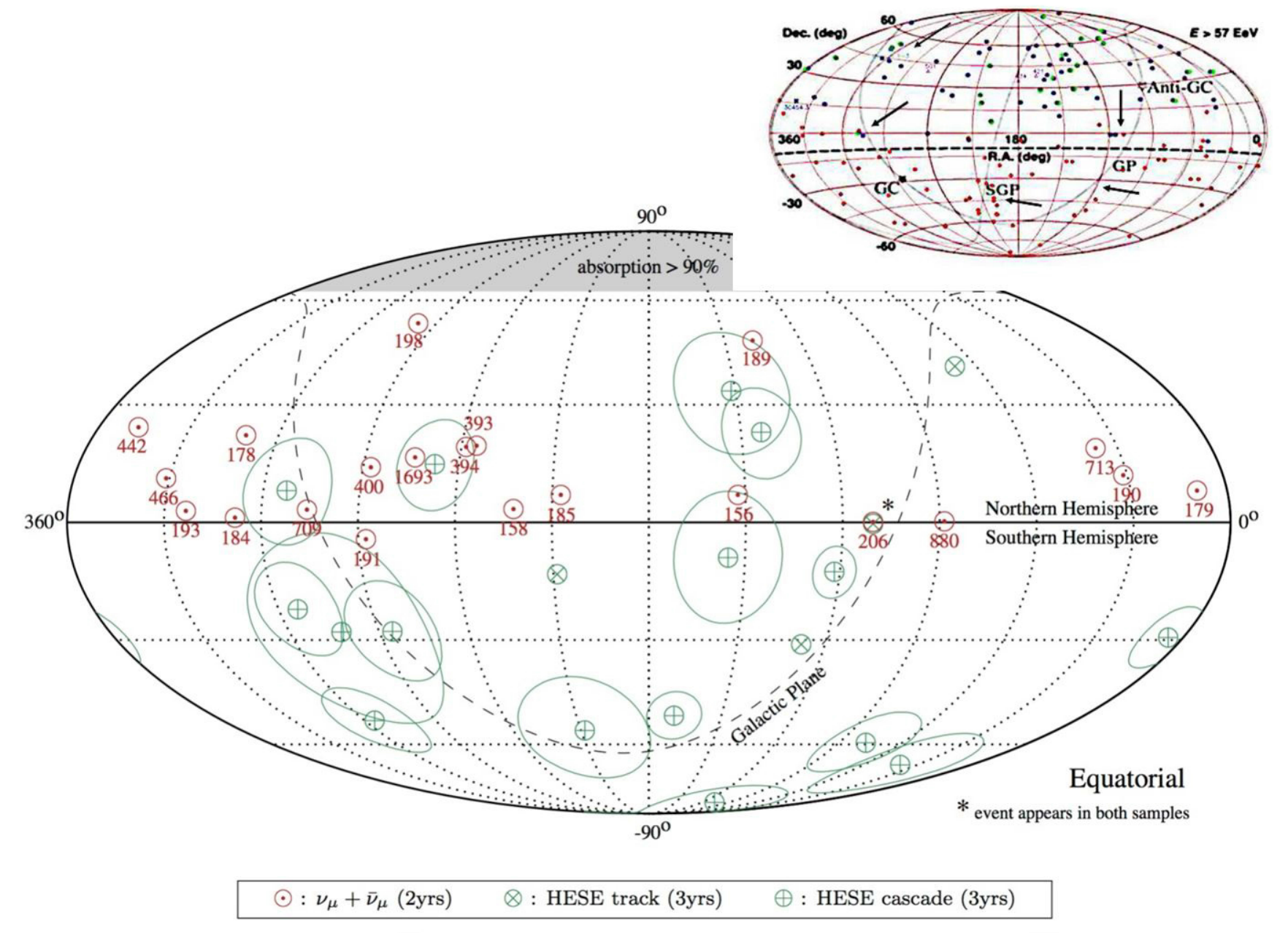}
    \caption{\label{fig3} Recent Crossing Muons in IceCube (21 events and 4 included ones) (2015) \cite{6e}, with an expected source distribution in an (arrowed) map considered  years earlier \cite{6e, 8a}. }\label{fig03}
\end{figure}

Indeed within such under sample of $\nu_\mu$ track (25), (four are UHE muon neutrino inside the detector, 21 are extra volume crossing  horizontal upward muons \cite{6e}),
 we found that two of the four labeled sources (SS433 and UHECR n.5) are located and overlaps with two of the new 21 muon tracks \cite{19}; see Fig. \ref{fig03}-\ref{fig04}.
A very fast view of the above events in Fig. \ref{fig3} show the presence of a large number of nearly horizontal  $\mu^{\mp}$ tracks whose
  whose origination is more probably of conventional $\pi^{\mp}, K^{\mp} $ nature,because along the
  horizons the $\pi^{\mp}$, K$^{\mp} $  have much longer ($\simeq 36$ times respect vertical ones) distances to decay. Moreover the flavor ratio for crossing (through going) muons as well as the absence of double bangs  favor a composition $\nu_{e}$, $\nu_{\mu}$, $\nu_{\tau}\simeq\frac{1}{2}$, $\frac{1}{2}$, 0 and not the astrophysical  expected one:  $\simeq\frac{1}{3}$, $\frac{1}{3}$, $\frac{1}{3}$. We concluded, from these early hints, that IceCube UHE events are still polluted, by a large fraction, $\simeq 70-90\%$ by atmospheric (conventional or prompt) neutrino noises \cite{19}.

The probability that by chance a new $\mu$ track would hit inside a spot that we had noted, within $\pm0.4^\circ$, is extremely low, below 5$\dot10^{-4}$ \cite{0c}. The eventuality that two out of the five of them would hit in such a narrow angle ($\lesssim1^\circ$) is  very improbable. Even in a more pessimistic source location  area, under $3^\circ\times3^\circ$ and within 21 new trials the probability that two events fall into two of the five narrow areas is extremely low: with
$p=5\cdot\frac{\pi\cdot3\cdot3}{1,5\cdot5\cdot10^4}\simeq10^{-2}
$, $n=21$, $k=2$, we find
$$
P=
\left(
\begin{array}{c}
21\\
2
\end{array}
\right)p^2q^{21}=1,73\%
$$
We must stress that, the way the narrow UHECR clustering around UHE energy muon track in ICECUBE, n.5, and the second crossing muon  overlapping (inside or through-going) is by itself an extremely rare event.  We didn't find any well known source in galactic plane along event $n.5$, but we wish to remind that almost half of the Fermi Gamma hardest sources are not identified.

\section{The neutrino muon  Astronomy}
\label{sec:nuastro}
\begin{figure}[t]
    \includegraphics[width=0.5\textwidth]{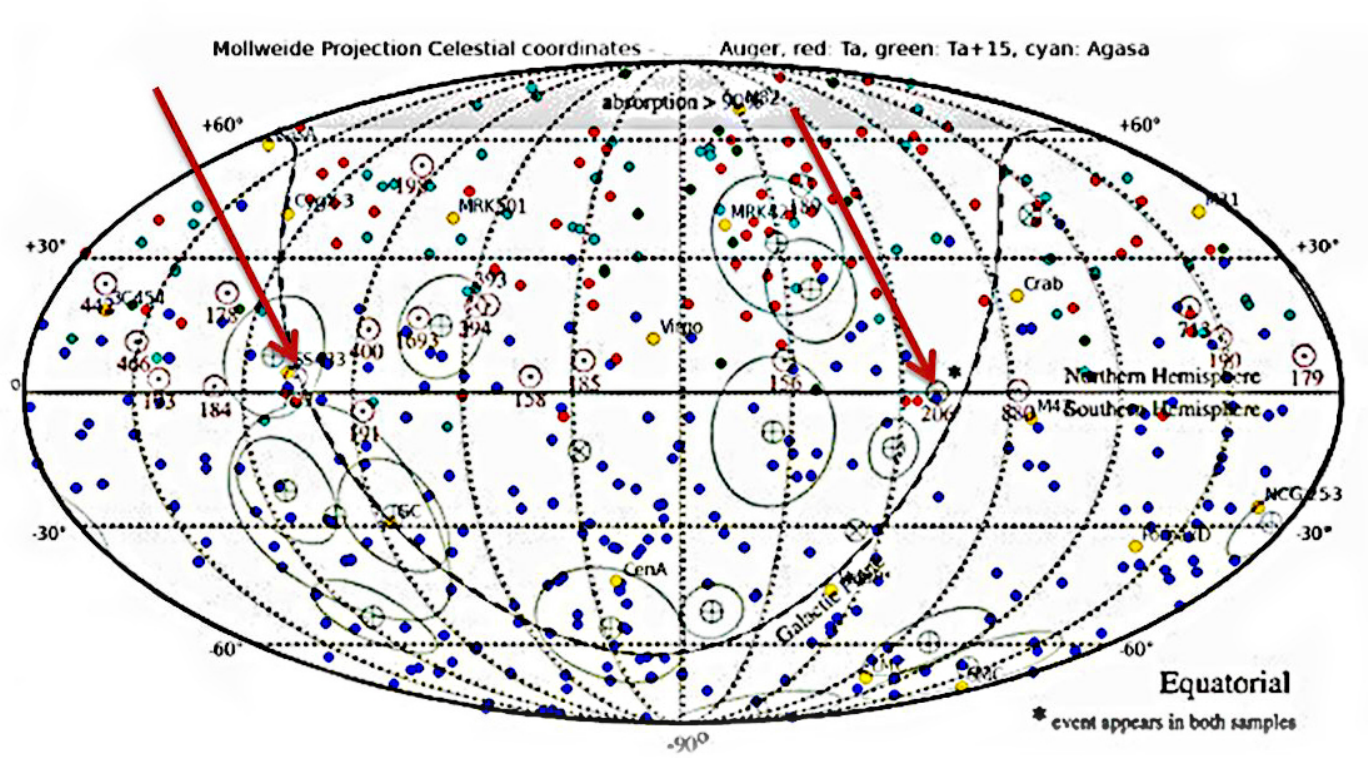}
    \caption{ Recent Crossing Muons in IceCube with overlap of the two main expected UHECR clustering shown by the red arrows.}\label{fig04}
\end{figure}

The apparent lack of clustering in UHE$\nu$ and UHECR events led most of the authors (Auger, TA, IceCube) to conclude that no sources are discovered yet and they postpone any interpretation to future increase in statistics \cite{1}.
 Moreover it was assumed (in contrast with our interpretation) that IceCube wide cone angle, the shower events, might be leading to a better sky correlation.
  We have to disagree: there are already enough hints and a first signs of an UHE $\nu_{\mu}$ overlap astronomy even if, we believe,
   a large fraction of ICECUBE are partially masked by prompt atmospheric neutrinos, \cite{19}.
We can summarize our view as follows:
\begin{enumerate}
\item Galactic UHE $\nu_{\mu}$ may correlate with rarest UHECR narrow clusterings.
\item UHECR smeared clustering are probably lightest nuclei (of extragalactic) nearby universe; they may be light nuclei or rarely, for galactic sources, heavy nuclei.
\item The smearing and bending of galactic fields are hiding the galactic center somehow repelling the UHECR on galactic plane boundary (the ridge).
\item A very few of the UHECR might correlate with UHE $\nu_{\mu}$ because, if they are galactic, their time travel is not too different (by a random walk respect) to a direct neutrino flight; a more far extragalactic source, even nearby as Cen A or M82,  maybe not be easily corresponding to an UHE$\nu$ activity today because the time lapse of flight along Mpc distances may differ respect transient activity:  short time scale AGN jet beaming, flashing and blazing may be occurring longer before.
\item The spread of  light nuclei-composed UHECR is very effective in smearing the direction information in our galaxy but the hardest and  first narrow clustering, as the one around SS433 and Cygnus X3 or event $n.5$ in IceCube muon track,  may be disentangled from the noisy UHECR sky.
\item Further more abundant muon tracks signals may soon offer more correlated overlapping within a narrow angle probing soon and fast their statistical weight.
\item While $\tau$ neutrino in IceCube are not yet observed, suggesting a prompt neutrino pollution, their upward $\tau$ air-showers from earth or mountain may shine in a very new and filtered $\nu$ astronomy \cite{0e1},\cite{0e3},\cite{20}. Near future
UHE $\nu$ $\tau$ double bang might anyway provide  IceCube with a more filtered astronomy, whose flavor ratio
might test our knowledge and expectation on neutrino mass mixing.
\end{enumerate}
\section{The tau airshower Astronomy}
The future $\tau$ airshower search  \cite{0e3}, through  via up going fluorescence shower lights in Auger  \cite{0f, 0g} and TA and  by \v{C}erenkov flashes, in ASHRA skies may offer a new signature of UHE$\nu$ astronomy. Furthermore this PeV neutrinos may also shine onto the clouds in Auger and TA  cloudy skies in an analogous way as described by us in \cite{20} for the MAGIC, VERITAS, HESS  telescope array. This can happen with sudden elliptical \v{C}erenkov ovals are reflected in nearby clouds with any telescope in the dark cloudy nights, if properly triggered. It should be note the possibility to reveal in AUGER or TA, as it is well known, by upward horizontal tau air-showers at EeVs energy; their upgoing showering within a few tens kilometer distance may hardly skim on the ground array (Cherenkov water elements), but they mostly shine by their lights on those  fluorescence telescope. Indeed few EeV tau airshower, even inclined  will start to decay usually one or more kilometer in atmosphere producing a diluted upward airshower with a negligible down ward skimming traces on ground.   Moreover also upward tens PeV tau airshower at much near (a few kilometer) distances from the fluorescence telescope might also be revealed (a few each year).
This possibility, while reducing or restricting the AUGER and TA effective area from thousands to tens kilometer size, it will offer a tool in a few PeVs energy neutrino windows astronomy to correlate to IceCube highest energy events. Naturally such nearest distance tau up-going of air showering require a different time trigger  (nearly tens time faster) and a wider and faster spread angle track.

Finally, the future GRAND radio array in China would be able to capture such up-going events if this arrays will be located along a deep long chain  mountain or at the top of mountain edges, like ASHRA experiment. This will guarantee a wider and certain solid angle acceptance, that is not possible while the detection is being located in an empty flat space. The use of existing aeolic top tower array in high mountain regions, may provide a best ideal (existing) structure array.

\section*{In memory}

This article is devoted to the memory of Martin Lewis Perl, who died last years on 30 September 2014, a Nobel physicist that gave life to the  unexpected  third (tau) lepton precursor opening the view to a new top and beauty quarks; we also commemorate, in analogy, the lost of lives this week,  of  top beauty girls, Hadar Cohen, Shlomit-Krigman, murdered at age 19 and 24, by unexpected stabbing.





\section*{References}
\bibliographystyle{elsarticle-num}



\end{document}